\documentclass[%
 aip,
 amsmath,amssymb,
 reprint,%
]{revtex4-1}

\usepackage{graphicx}
\usepackage{dcolumn}
\usepackage{bm}

\usepackage[utf8]{inputenc}
\usepackage[T1]{fontenc}
\usepackage{mathptmx}

\usepackage[pdftex]{color}

\begin{document}


\title{Trends in elastic properties of Ti-Ta alloys from first-principles calculations}

\author{Tanmoy Chakraborty}
\email{tchakra7@jhu.edu.}
\affiliation{ 
Interdisciplinary Centre for Advanced Materials Simulation, Ruhr-Universit{\"a}t Bochum, 44780 Bochum, Germany
}
\affiliation{Department of Materials Science and Engineering, Johns Hopkins University, Baltimore, Maryland 21218, USA}
\author{Jutta Rogal}%
\affiliation{ 
Interdisciplinary Centre for Advanced Materials Simulation, Ruhr-Universit{\"a}t Bochum, 44780 Bochum, Germany
}%

\date{\today}

\begin{abstract}
The martensitic start temperature ($M_{\text{s}}$) is a technologically fundamental characteristic of high-temperature shape memory alloys. We have recently shown [Phys. Rev. B 94, 224104 (2016)] that the two key features in describing the composition dependence of $M_\text{s}$ are the $T=0$~K phase stability and the difference in vibrational entropy which, within the Debye model, is directly linked to the elastic properties.
Here, we use density functional theory together with special quasi-random structures to study the elastic properties of disordered martensite and austenite Ti-Ta alloys as a function of composition. We observe a softening in the tetragonal shear elastic constant of the austenite phase at low Ta content and a \emph{non-linear} behavior in the shear elastic constant of the martensite. A minimum of 12.5\% Ta is required to stabilize the austenite phase at $T = 0$~K. Further, the shear elastic constants and Young's modulus of martensite exhibit a maximum for Ta concentrations close to 30\%. Phenomenological, elastic-constant-based criteria suggest that the addition of Ta enhances the strength, but reduces the ductile character of the alloys. In addition, the directional elastic stiffness, calculated for both martensite and austenite, becomes more isotropic with increasing Ta content. The reported trends in elastic properties as a function of composition may serve as a guide in the design of alloys with optimized properties in this interesting class of materials.
\end{abstract}

\maketitle

\section{\label{sec:intro}Introduction}

Ti-Ta alloys are among the most promising candidates for high-temperature shape memory alloys (HTSMAs), materials which exhibit martensitic phase transformations to reversely convert heat into mechanical strain and that are currently being employed as, e.g., sensors or actuators in automotive and aerospace applications~\cite{Ma_imr, Pio1, Firstov}. Among various types of shape memory alloys such as Ti-Mo~\cite{Miyazaki-438, Kim-54,Kim-45} and Ti-Nb~\cite{Matsumoto-465, Inamura-253}, Ti-Ta solid solutions not only present advantages including relatively high transformation temperatures ($M_{\text{s}}$ $>$ 373 K) and a low tendency to form the hexagonal $\omega$ phase, which is detrimental for the shape memory effect and causes embrittlement~\cite{Pio1}, but also show excellent cold workability~\cite{Pio2}. Theoretical and experimental investigations on Ti-Ta and Ti-Ta-based alloys demonstrated that the martensitic transformation temperature $M_{\text{s}}$, as well as other physical properties of the two phases are strongly composition dependent~\cite{chakraborty_prb, Pio1, Tanmoy, Ferrari_SMSE, Ferrari_PRM2019}.

Recently, we have shown that the relative phase stability at $T=0$~K and the  difference in the vibrational free energy contribution (which within the Debye model is derived from the elastic properties) between the martensite (orthorhombic, $\alpha''$, low-temperature phase) and the austenite (cubic, $\beta$, high-temperature phase)  phases in Ti-Ta are the two critical parameters that govern the composition dependence of $M_{\text{s}}$~\cite{chakraborty_prb}. The relative stability of the different competing phases in Ti-Ta alloys has been studied extensively in our previous work~\cite{Tanmoy}. 
Wu \emph{et al.}~\cite{Wu_12} studied the phase stability and also elastic properties, but of only the $\beta$ phase in Ti-Ta alloys
and without accounting for chemically disordered structures. A careful analysis of changes in the elastic properties of both, austenite and martensite, upon varying the Ta concentration can improve our understanding of the chemical interplay that may affect the martensitic transformation in shape memory alloys as well as provide useful insights into designing HTSMAs with desired properties. 
Moreover, trends in elastic properties can be used to determine the conditions of mechanical stability in cubic Ti-based alloys~\cite{Skripnyak} and qualitatively assess the alloy mechanical behavior. 
A comprehensive investigation of the elastic properties, in particular of the martensite phase, is, to the best of our knowledge, still missing in theory and experiment, motivating the present study. 

Here, we use density functional theory (DFT) to investigate the single-crystal elastic constants and derived polycrystalline properties of the martensite and austenite phases of Ti-Ta alloys as a function of composition. Our 
results show that the tetragonal shear elastic constant of the austenite phase softens at very low Ta content and then gradually stiffens upon further alloying with Ta. This indicates that the mechanical stability of the austenite increases with increasing Ta content. The shear elastic constants of the martensite phase exhibit a \emph{non-linear} behavior with a maximum value at intermediate Ta concentrations. The polycrystalline properties of the two phases derived from the single crystal elastic constants correspondingly reveal similar trends.
Based on the empirical criteria proposed by Pugh~\cite{Pugh}, Pettifor~\cite{Pettifor}, and most recently by Niu \emph{et al.}~\cite{Niu}, our results suggest that Ta enhances the strength of Ti-Ta binary alloys and lowers the ductility. 
We have also investigated the effect of Ta alloying on the anisotropy of both phases by calculating different anisotropy indices and directional-dependent Young's moduli. The anisotropy of both phases decreases smoothly with increasing Ta content. 

The paper is organized as follows: In Section~\ref{method} we describe the computational details and the  Voight-Reuss-Hill (VRH) approximation for polycrystalline elastic moduli. In Section~\ref{elastic properties} we present our \emph{ab initio} results on the single crystal elastic and polycrystalline properties, as well as the elastic anisotropy. We conclude our findings in Section~\ref{conclusions}.

\section{Computational details}\label{method}

First principles density functional theory (DFT) calculations were carried out using the Vienna $Ab$ $initio$ Simulation Package (VASP)~\cite{Kresse_15, Kresse_11169, Kresse_1758}. We use Perdew-Burke-Ernzerhof 
gradient corrected exchange-correlation functional (PBE-GGA)~\cite{Perdew_3865} for all calculations. Projector augmented wave (PAW)~\cite{Kresse_1758, Bloechl} potentials were used for all calculations  including 
3$p$ and 5$p$ electrons for Ti and Ta in the valence shell, respectively. A cutoff energy of 300 eV for the plane waves and Methfessel-Paxton scheme was used to integrate the Brillouin zone (BZ) with a smearing
of $\sigma$ = 0.05 eV. 
With this setup,  total energies are converged to within 4 meV/atom. We have used special quasi-random structure (SQS) method~\cite{Zunger} as implemented in the modified version~\cite{Pezold, Kossmann} of the ATAT package~\cite{Walle} to model the chemical disorder of the martensite and austenite phases.
We considered six different compositions  with 12.5\%, 18.75\%, 25\%, 31.25\%, 37.5\% and 43.75\% Ta 
(details concerning the corresponding SQS supercells can be found in the Supplementary Material (SM)).

The elastic constants of the martensite and austenite phases were calculated using the stress-strain method as implemented in VASP~\cite{Page}. The cubic austenite and  orthorhombic martensite phases have three ($C_{11}$, $C_{12}$, and $C_{44}$) and nine ($C_{11}$, $C_{22}$, $C_{33}$, $C_{44}$, $C_{55}$, $C_{66}$, $C_{12}$, $C_{13}$ and $C_{23}$) independent elastic constants, respectively. For cubic systems, $C_{11}$ represents the uniaxial deformation along the [001] direction, $C_{12}$ is the shear stress at the (110) crystal plane along the [110] direction, and $C_{44}$ represents a shear deformation on the (100) crystal plane.
The elastic tensor matrix for the martensite (orthorhombic) phase is given by
\[ \left( \begin{array}{cccccc}
C_{11} & C_{12} & C_{13} & 0 & 0 & 0 \\
C_{12} & C_{22} & C_{23} & 0 & 0 & 0 \\
C_{13} & C_{22} & C_{33} & 0 & 0 & 0 \\
0 & 0 & 0 & C_{44} & 0 & 0 \\
0 & 0 & 0 & 0 & C_{55} & 0 \\
0 & 0 & 0 & 0 & 0 & C_{66} \end{array} \right)\]
For the austenite (cubic), $C_{11}=C_{22}=C_{33}$, $C_{12}=C_{13}=C_{23}$, and $C_{44}=C_{55}=C_{66}$. Using SQSs to describe the chemical disorder results in a lowering of the symmetry and minor differences in the elastic constants that are equivalent in the cubic phase.  Correspondingly, the elastic constants for the $\beta$ phase have been averaged as~\cite{Ferenc, Gao_2013}
\begin{align}
\overline{C_{11}} (\beta) = \frac{(C_{11}+C_{22}+C_{33})}{3}\bigg |_{\text{SQS}}  \quad , \\ \nonumber
\overline{C_{12}} (\beta) = \frac{(C_{12}+C_{13}+C_{23})}{3}\bigg |_{\text{SQS}} \quad , \\  \nonumber
\overline{C_{44}} (\beta) = \frac{(C_{44}+C_{55}+C_{66})}{3}\bigg |_{\text{SQS}} \quad . \\  \nonumber
\end{align}

The polycrystalline elastic moduli of the martensite and austenite phases have been computed within the Voigt-Reuss-Hill (VRH) approximation~\cite{Voigt,Reuss,Hill}. The Voigt approximation~\cite{Voigt} provides an upper bound of the bulk ($B$) and shear ($G$) modulus, whereas the Reuss approximation~\cite{Reuss} represents a lower bound of $B$ and $G$. The relationship between the elastic constants matrix \textbf{C} and the compliance matrix \textbf{S} elements with the polycrystalline bulk and shear moduli for the martensite and the austenite are given by~\cite{Xiao_109, Connetable_79}:

\begin{itemize}
 \item Austenite phase:
\end{itemize}

\begin{equation}
 B_V = B_R = (C_{11} + 2C_{12})/3,
\end{equation}

\begin{equation}
 G_V = (C_{11} - C_{12} +3C_{44})/5,
\end{equation}

\begin{equation}
 G_R = 5(C_{11} - C_{12})C_{44}/[4C_{44} + 3(C_{11} - C_{12})],
\end{equation}

\vspace{1em}
\begin{itemize}
 \item Martensite phase:
\end{itemize}

\begin{equation}
 B_V = \frac{1}{9}(C_{11} + C_{22} + C_{33}) + \frac{2}{9}(C_{12} + C_{23} + C_{13})
\end{equation}

\begin{multline}
 G_V = \frac{1}{15}[C_{11} + C_{22} + C_{33} - (C_{12} + C_{23} + C_{13})] \\ + \frac{1}{5} (C_{44} + C_{55} + C_{66})
 \end{multline}

\begin{equation}
 \frac{1}{B_R} = (S_{11} + S_{22} + S_{33}) + 2(S_{12} + S_{13} + S_{23})
\end{equation}

\begin{align}
 \frac{1}{G_R} = \frac{1}{15}[4(S_{11} + S_{22} + S_{33}) - 4(S_{12} + S_{13} + S_{23}) \\  \nonumber
 + 3(S_{44} + S_{55} + S_{66})]
\end{align}
where $S_{11}$, $S_{12}$, etc. are the elements of the elastic compliance matrix with \textbf{S} = \textbf{C}$^{-1}$. $B_V$, $B_R$, $G_V$, and $G_R$ are the bulk modulus and shear modulus from the Voigt and Reuss equations, respectively. 
Polycrystalline elastic properties such as $B$ and $G$ can be calculated from the elastic constants using the VRH approximation 
\begin{equation}
 B = \frac{1}{2}(B_V + B_R)
\end{equation}
\begin{equation}
 G = \frac{1}{2}(G_V + G_R) \quad .
\end{equation}
Furthermore, the Young's modulus ($E$) within the VRH framework is given by 
\begin{equation}
 E = \frac{9BG}{3B + G} \quad .
\end{equation}


\section{Results}\label{elastic properties}
\subsection{Composition dependent elastic constants}\label{Composition dependent elastic constants}
\subsubsection{Single crystal elastic constants}

In this section, we report the trends in elastic constants of the austenite and martensite as a function of composition. The elastic constants for pure bcc Ti and bcc Ta (see Table S1 in the Supplementary Material) are in good agreement with previous \emph{ab initio} and experimental results~\cite{Ikehata, Allard}. According to the Born-Huang criteria~\cite{Zhi} a cubic phase is mechanically stable if
\begin{equation}
 C_{11} > 0, \; C_{44} > 0, \; C_{11} > |C_{12}|, \; (C_{11}+2C_{12}) > 0 \quad .
\end{equation}
\begin{figure}
\begin{center}
 \includegraphics[width=0.45\textwidth]{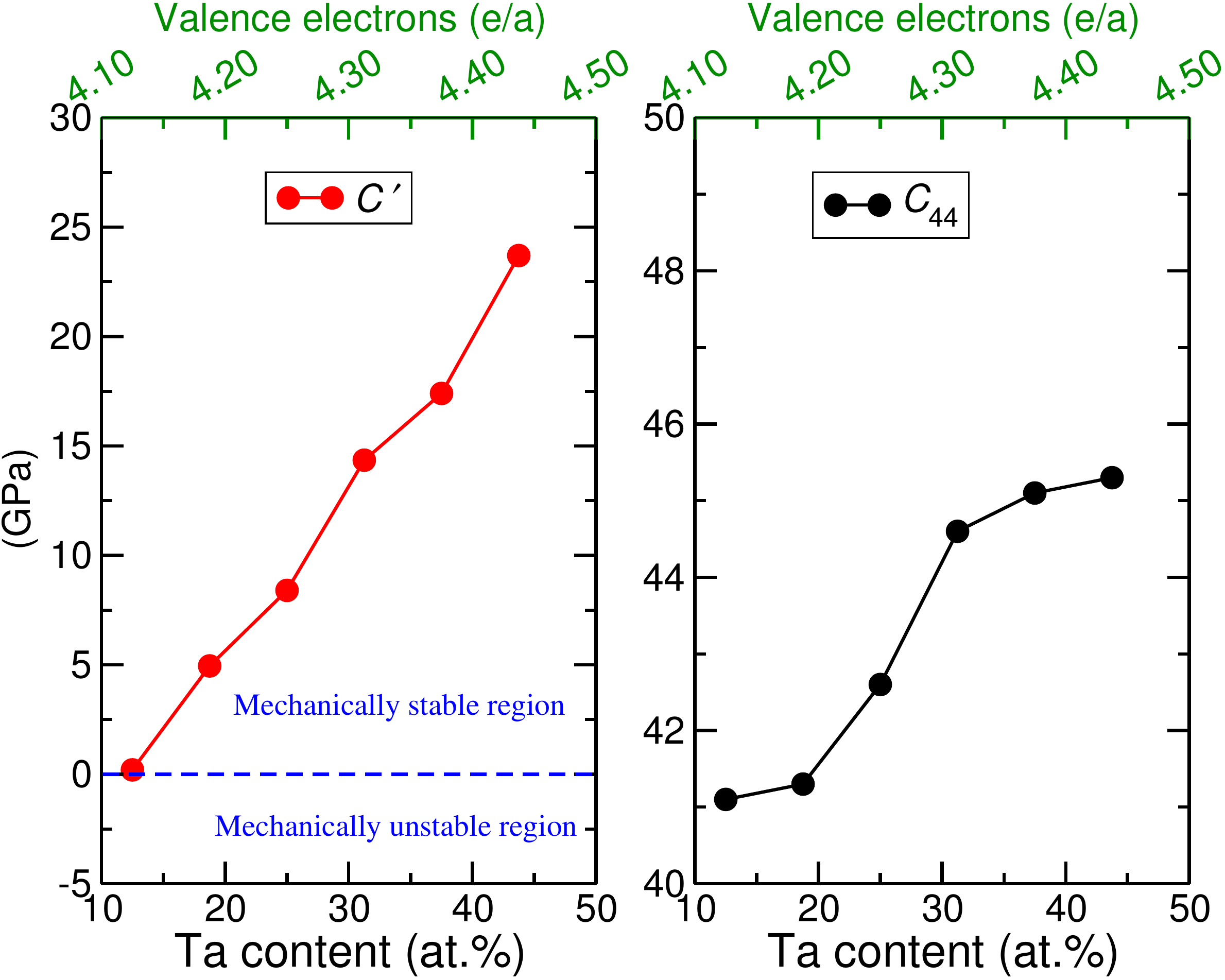}
\caption{\label{fig-1}
Tetragonal shear constant ($C'$) and trigonal shear constant ($C_{44}$) of the austenite phase as a function of Ta content.
}
\end{center}
\end{figure}
The austenite phase in Ti-Ta satisfies the Born-Huang stability criteria for all compositions studied, 
that is the austenite phase appears to be 
elastically stable (values are listed in Table~S2 of the Supplementary Material). The tetragonal shear constant ($C' = (C_{11} - C_{12})$/2) is strongly dependent on composition, increasing with increasing  Ta content as shown in Fig.~\ref{fig-1}. 
This implies that alloying Ta stabilizes the bcc austenite phase, which is expected as Ta is a $\beta$-stabilizer. Below 12.5\% Ta our calculations predict that the austenite phase will be mechanically unstable at $T = 0$~K as $C'$ becomes negative, which is in agreement with the fact that bcc Ti is mechanically and dynamically unstable at $T$ = 0 K~\cite{Mei_80, Hu_107}. 
The top $x$-axis of Fig.~\ref{fig-1} displays the corresponding change in the $e/a$ ratio (the number of electrons per atom).
The monotonic correlation between $C'$ and $e/a$ has been identified experimentally by Fischer \emph{et al.}~\cite{Fischer} for bcc transition metals, and has also been recognized later theoretically for $\beta$-type Ti alloys such as Ti-Nb~\cite{Hu-93, Lai}, consistent with our findings for Ti-Ta. Similar results have also been reported for Ti-V solid solutions~\cite{Skripnyak}, where $\approx 18$~\% V are required to mechanically stabilize the bcc phase in Ti-V alloys.

In the right graph of Fig.~\ref{fig-1}, the dependence of the trigonal shear constant ($C_{44}$) on composition is shown. $C_{44}$ also increases with increasing Ta concentration, but is much less sensitive to a change in composition (increase by only $\approx$~10\%) compared to $C'$. Similar findings have also been reported for Ti-Nb~\cite{Lai, Moreno} and Ti-V~\cite{Skripnyak}, where $C_{44}$ is also nearly independent of Nb and V alloying, respectively. 

The mechanical stability criteria for the orthorhombic martensite phase are~\cite{Zhi}
\begin{multline}
\label{mart_eq}
 C_{11}>0, \; C_{22}>0, \; C_{33}>0, \; C_{44}>0, \; C_{55}>0, \; C_{66}>0, \\ 
 [C_{11}+C_{22}+C_{33}+2(C_{12}+C_{13}+C_{23})]>0, \\ 
 (C_{11}+C_{22}-2C_{12})>0, \; (C_{11}+C_{33}-2C_{13})>0, \\
 (C_{22}+C_{33}-2C_{23})>0 \quad .
\end{multline}
The elastic constants of the martensite are shown in Fig.~\ref{fig:fig-2} as a function of Ta concentration (values are provided in Table~S3 of the Supplementary Material). All criteria in Eq.~\eqref{mart_eq} are satisfied implying the existence of a mechanically stable martensite phase over the entire composition range studied in this work. 
The elastic constants, apart from characterizing the mechanical and dynamical stability of materials, can often also be related to the crystallographic axes to obtain insights into the incompressibility along different directions. $C_{11}$, for example, characterizes resistance to linear compression under uniaxial stress along the $x$ direction~\cite{Huihui, Ganeshraj} which also represents atomic bonding along the $x$ direction~\cite{Masys}. Between the austenite and martensite the orientation relationship (OR)  is given by
\begin{align}
\label{OR1}
 [110]_{\text{aus}}~~||~~[100]_{\text{mar}} \quad , \\ \nonumber
 [\overline{1}10]_{\text{aus}}~~||~~[010]_{\text{mar}} \quad , \\ \nonumber
 [001]_{\text{aus}}~~||~~[001]_{\text{mar}} \quad.
 \end{align}
To be consistent with common notations used in experiments for the martensite phase with lattice vectors $a$ < $c$ < $b$, we have setup our supercells such that 
\begin{align}
\label{OR2}
a = x = [001]_{\text{aus}}~~||~~[100]_{\text{mar}} \quad , \\ \nonumber
b = y = [110]_{\text{aus}}~~||~~[010]_{\text{mar}} \quad , \\ \nonumber
c = z = [\overline{1}10]_{\text{aus}}~~||~~[001]_{\text{mar}} \quad.
 \end{align}
For low Ta content, $C_{22} \approx C_{33} > C_{11}$, indicative of the fact that the martensite phase reveals the largest incompressibility along the $b$ and $c$ axis, and that it is relatively compressible along the $a$ axis. With increasing Ta content, $C_{11}$ increases significantly, while $C_{22}$ and  $C_{33}$ exhibit smaller changes.  $C_{11}$  is very similar in the martensite and austenite (open circles in Fig.~\ref{fig:fig-2}) phases, which is reasonable since considering the OR between the two phases $C_{11}$ corresponds to a deformation in the same crystallographic direction.
Also, wee see that at 43.75\% Ta, the value of $C_{11}$ for the austenite phase becomes higher than that of the martensite phase, indicative of a higher mechanical stability of the austenite phase at high Ta content.

\begin{figure}
\begin{center}
 \includegraphics[width=0.48\textwidth]{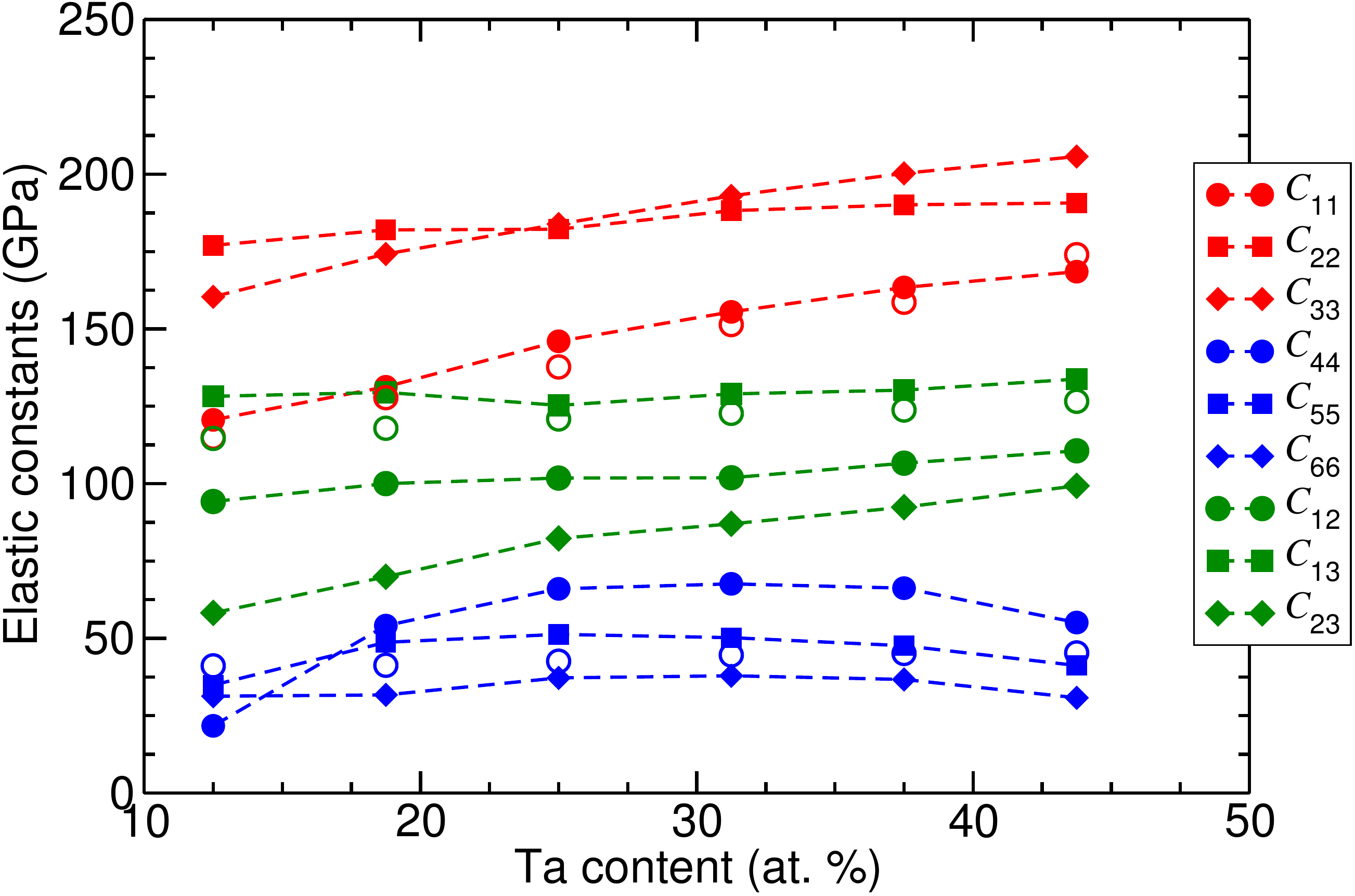}
 \caption{\label{fig:fig-2}
The nine elastic constants of the orthorhombic martensite phase are shown by filled symbols as a function of Ta content. The primary, shear, and off-diagonal elastic constants are shown in red, blue, and dark green, respectively. $C_{11}$, $C_{12}$, $C_{44}$ of the cubic austenite phase are also shown with open circles in red, blue, and dark green, respectively.
}
\end{center}
\end{figure}

The shear elastic constants $C_{44}$, $C_{55}$, and $C_{66}$ in  Fig.~\ref{fig:fig-2} show approximately the same value at 12.5\% Ta. The similarities in shear moduli at low Ta content might be indicative of the tendency of Ti-Ta to transform into the energetically more stable hexagonal $\omega$ phase~\cite{Tanmoy}, for which $C_{44}$ = $C_{55}$ = $C_{66}$. For larger Ta concentrations, the shear moduli of the martensite maintain the trend $C_{44}$ $>$ $C_{55}$ $>$ $C_{66}$ over the entire investigated composition range. This indicates that the martensite phase possesses the greatest shear resistance within the (100) plane and that the softest shearing transformation is observed along the (001) plane. 
While $C_{55}$ and  $C_{66}$ change only slightly, $C_{44}$ exhibits a non-linear behaviours with an increase up to about 30\% Ta, followed by a slight decrease.

At high Ta content, there seems to be an interesting pattern where the primary elastic constants ($C_{11}$, $C_{22}$, and $C_{33}$), off-diagonal elastic constants ($C_{12}$, $C_{13}$, and $C_{23}$), and the shear constants ($C_{44}$, $C_{55}$, and $C_{66}$) tend to converge to three distinct sets of values, respectively (i.e., to say, $C_{11}$, $C_{12}$, $C_{44}$), cf. Fig.~\ref{fig:fig-2}. One could imagine that these three sets qualitatively represent the three elastic constants of a cubic phase. This implies that the martensite is on the verge of martensite $\to$ austenite phase transformation, which is in good agreement with previous theoretical and experimental findings~\cite{Tanmoy, Dobromyslov_44, Dobromyslov_438, Zhou283}. 
It is further supported by the fact that both, the $b/a$ and $c/a$, ratios of the martensite decrease with increasing Ta content~\cite{chakraborty_prb}, making the martensite phase more cubic.

\begin{figure}
\begin{center}
 \includegraphics[height=0.4\textwidth]{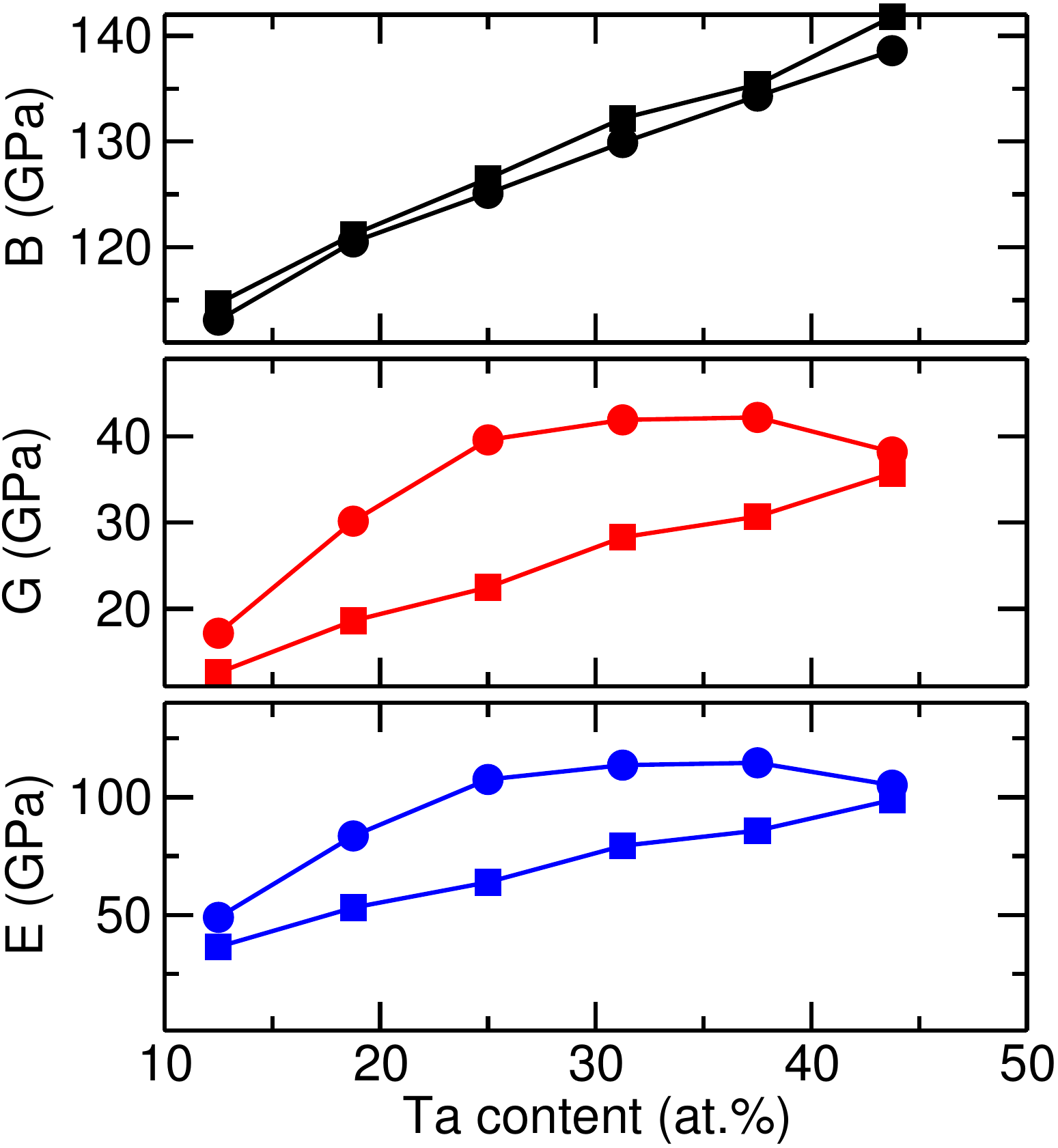}
\caption{\label{fig:fig-3}
Trends in the bulk modulus (B), shear modulus (G) and Young's modulus (E) of the martensite ($\alpha''$, circle) and austenite ($\beta$, square) as a function of Ta content.}
\end{center}
\end{figure}


\subsubsection{Polycrystalline properties}

Using the values of elastic constants together with VRH approximation we studied various polycrystalline properties of the martensite and the austenite phase as a function of composition. 
The corresponding values for the bulk ($B$), shear ($G$), and Young's ($E$) moduli are shown in Fig.~\ref{fig:fig-3}. The bulk modulus (top graph) increases linearly with increasing Ta content for both phases, and appears to be a simple linear interpolation between the bulk moduli of pure Ti and Ta. The shear modulus mainly depends upon the shear strain and is a measure of the stability of a phase upon shear. As shown in the middle panel of Fig.~\ref{fig:fig-3}, $G$ also increases linearly with Ta content for the cubic austenite phase. For the martensite phase, however, $G$ initially increases linearly up to $\sim$25\% Ta, then exhibits a plateau-like region up to $\sim$40\% Ta, before it decreases upon further addition of Ta. This interesting behavior of $G$ for the martensite phase originates in the corresponding shear elastic constants ($C_{44}$, $C_{55}$, and $C_{66}$) that show a similar trend, see Fig.~\ref{fig:fig-2}. It is to be noted that although the value of $G$ differs maximum among both the phases within the 25\%-35\% Ta content regime, at very low and high Ta content the values tend to converge, which might be related to the expected phase transformations ($\alpha''$, $\beta$) $\to$ $\omega$ and $\alpha''$ $\to$ $\beta$ at low and high Ta content, respectively. The Young's modulus reflects the stiffness of a material which in turn is connected to the bonding strength within the material. The dependence of $E$ on the composition is shown in the lower panel of Fig.~\ref{fig:fig-3}.
The change in $E$ with Ta alloying is similar to that in $G$ for both  phases, that is linear for the austenite and non-linear with a clear maximum for the martensite phase. 
The trends in $B$, $G$, and $E$ of the austenite and martensite phase are in very similar to that observed for Ti-Nb alloys~\cite{Moreno, Lai}.

\begin{figure}
\begin{center}
  \includegraphics[height=0.32\textwidth]{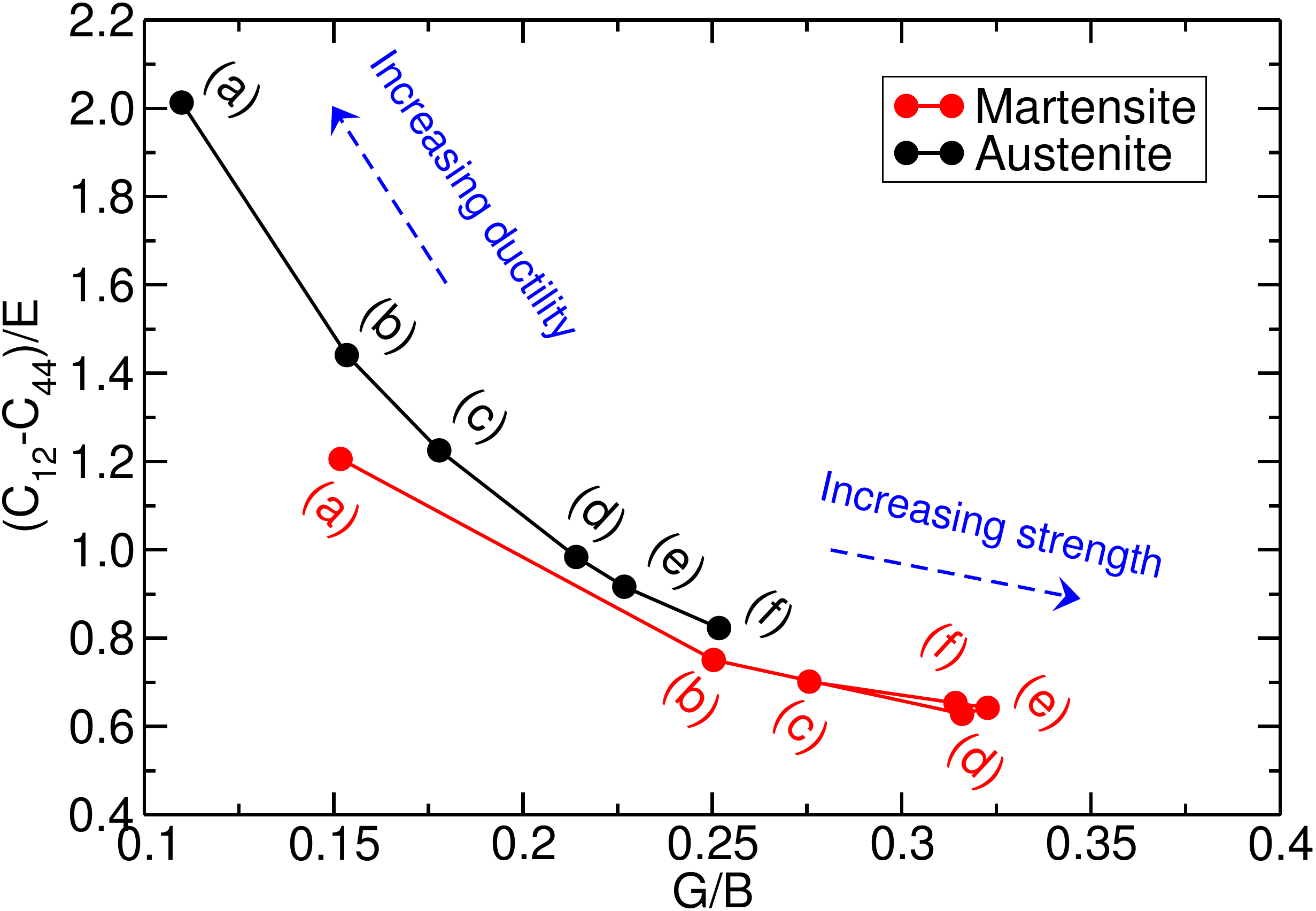}
\caption{\label{fig:fig-4}
A qualitative strength-ductile phase diagram of the martensite and the austenite as a function of Ta alloying. (a)-(f) points represent 12.5\%, 18.75\%, 25\%, 31.25\%, 37.5\% and 43.75\% Ta content respectively.
}
\end{center}
\end{figure}

To investigate the brittle-ductile properties in these alloys, we follow a more recently discovered universal ductile-to-brittle criterion reported by Niu~\emph{et al.}~\cite{Niu}, which is essentially a renormalized hyperbolic correlation derived by dividing the Cauchy pressure by the Young's modulus (revised Cauchy pressure) and plotting it as a function of Pugh's ratio, $G/B$. According to Niu~\emph{et al.}~\cite{Niu}, the higher Pugh's ratio the higher the material's  strength and consequently hardness. On the other hand, a higher value of the revised Cauchy pressure indicates a more ductile material. In Fig.~\ref{fig:fig-4}, the corresponding results are shown for the martensite and austenite phases. With increasing  Ta content both, the martensite and austenite, tend to increase strength i.e, loose their ductility. However, we also note that due to the \emph{non-linear} elastic response of the martensite it becomes more ductile again
upon further alloying Ta  above $\sim$40\%.  There is currently no existing experimental evidence to support our brittle-ductile findings, however, we note that $\beta$ stabilizing element, in general, tends to enhances the Pugh's ratio e.g, in Ti-Nb~\cite{Lai}, 
similar to our findings for Ti-Ta.


\subsection{Elastic anisotropy}\label{directional dependence}

Elastic anisotropy plays a crucial role in the formation of microcracks in materials~\cite{Tvergaard} as well as significantly impacts the mechanical properties of materials~\cite{tasnadi-apl}. 
It is therefore important to study elastic anisotropy in materials to understand their durability that, in turn, can assist the design of improved materials. 
In this work, we have computed two important anisotropy indices: (1) Universal elastic anisotropy ($A^{U}$)~\cite{Ranganathan_prl}, which is also interpreted as a generalization of the Zener anisotropy index, and (2) Chung-Buessem percent shear anisotropy ($A^C$)~\cite{Chung}. 
These are defined as
\begin{figure}
\begin{center}
 \includegraphics[width=0.48\textwidth]{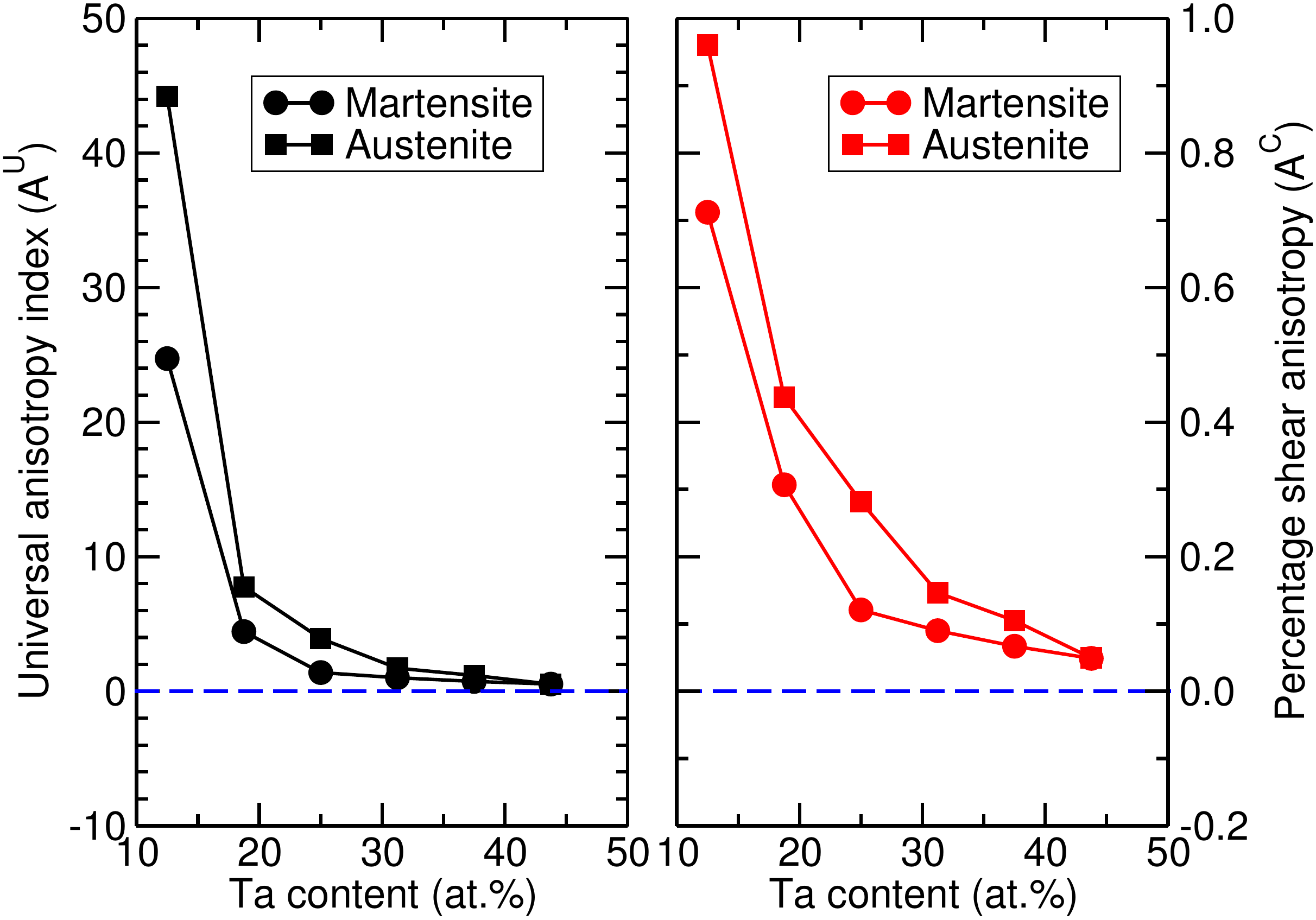}
 \caption{\label{fig:fig-5}
 The universal anisotropy index ($A^{U}$) and the Chung-Buessem anisotropy index ($A^C$) of the martensite (circle) and austenite (square) as a function of Ta content.
}
\end{center}
\end{figure}
\begin{figure*}
    \centering
        \includegraphics[width=0.18\textwidth]{mart-1875.pdf}
        \includegraphics[width=0.18\textwidth]{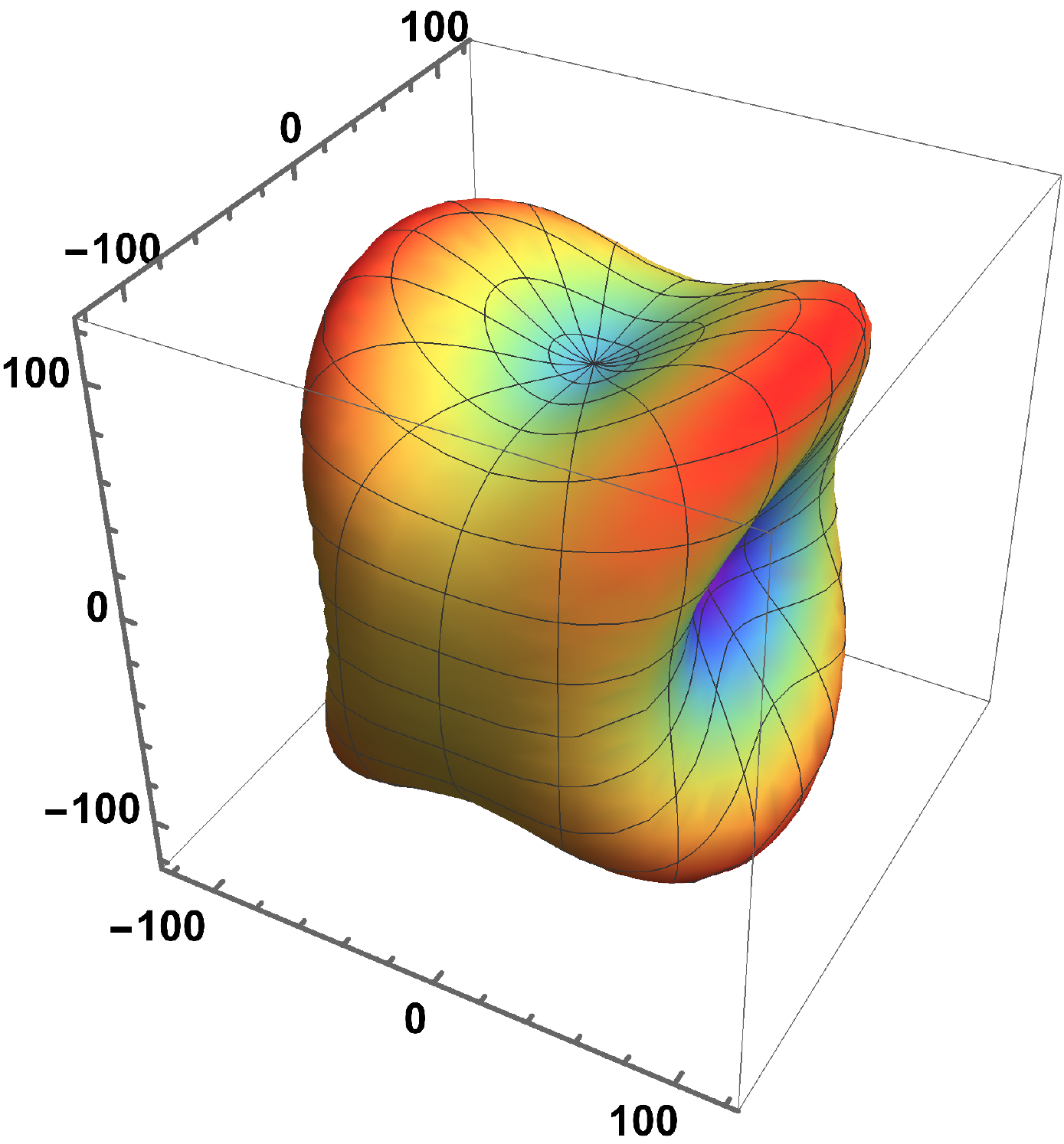}
        \includegraphics[width=0.18\textwidth]{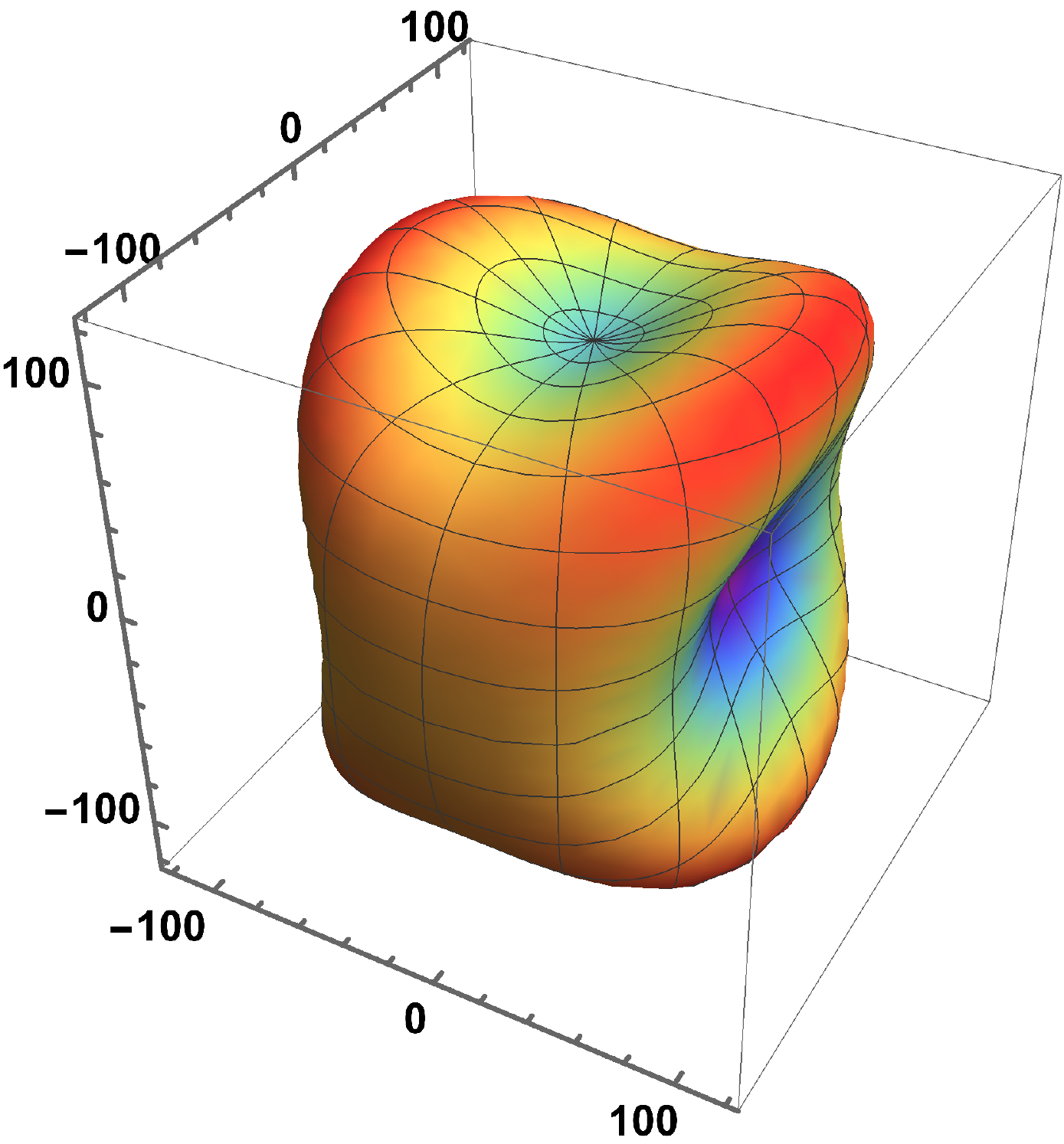}
        \includegraphics[width=0.18\textwidth]{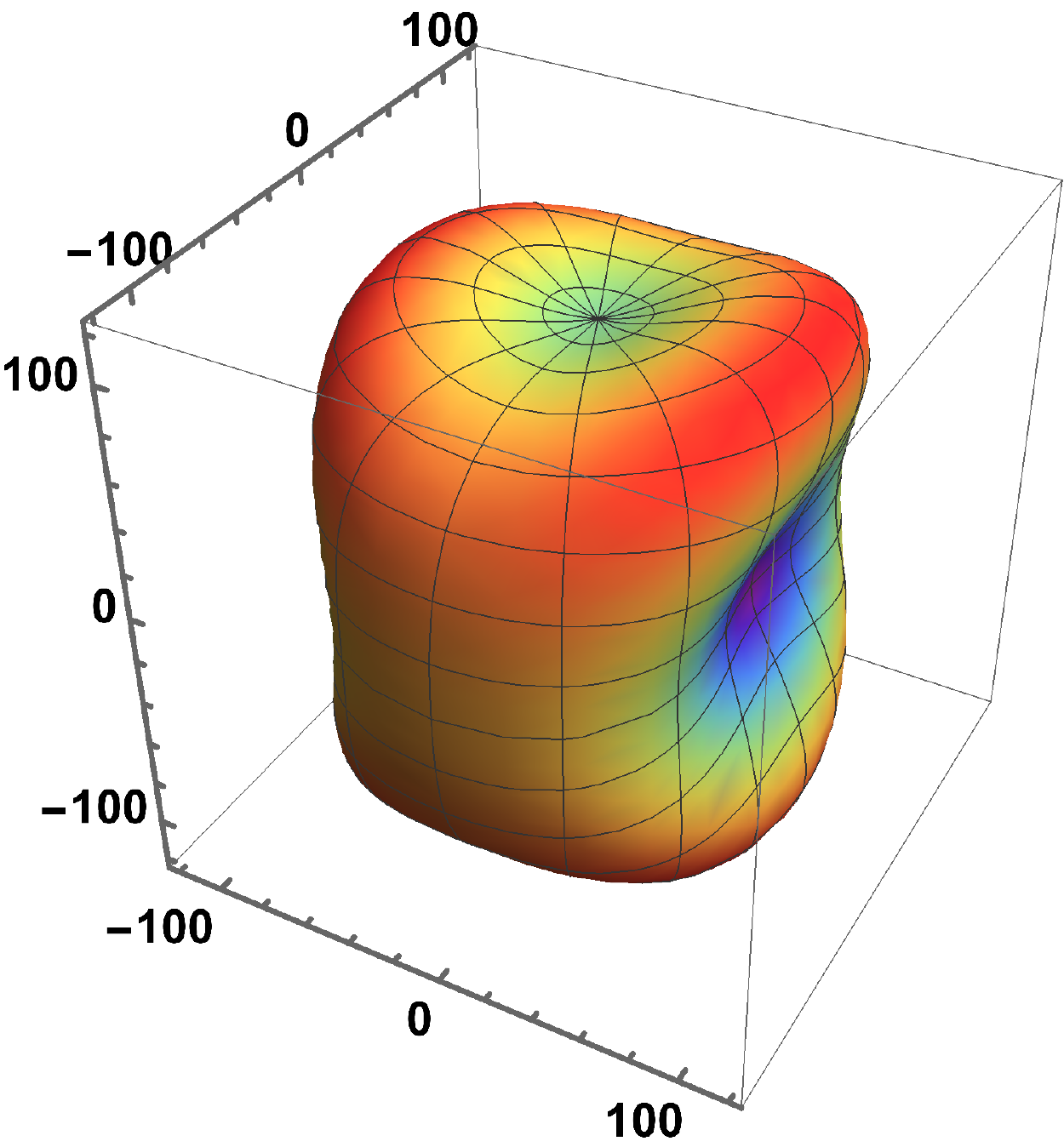}
        \includegraphics[width=0.18\textwidth]{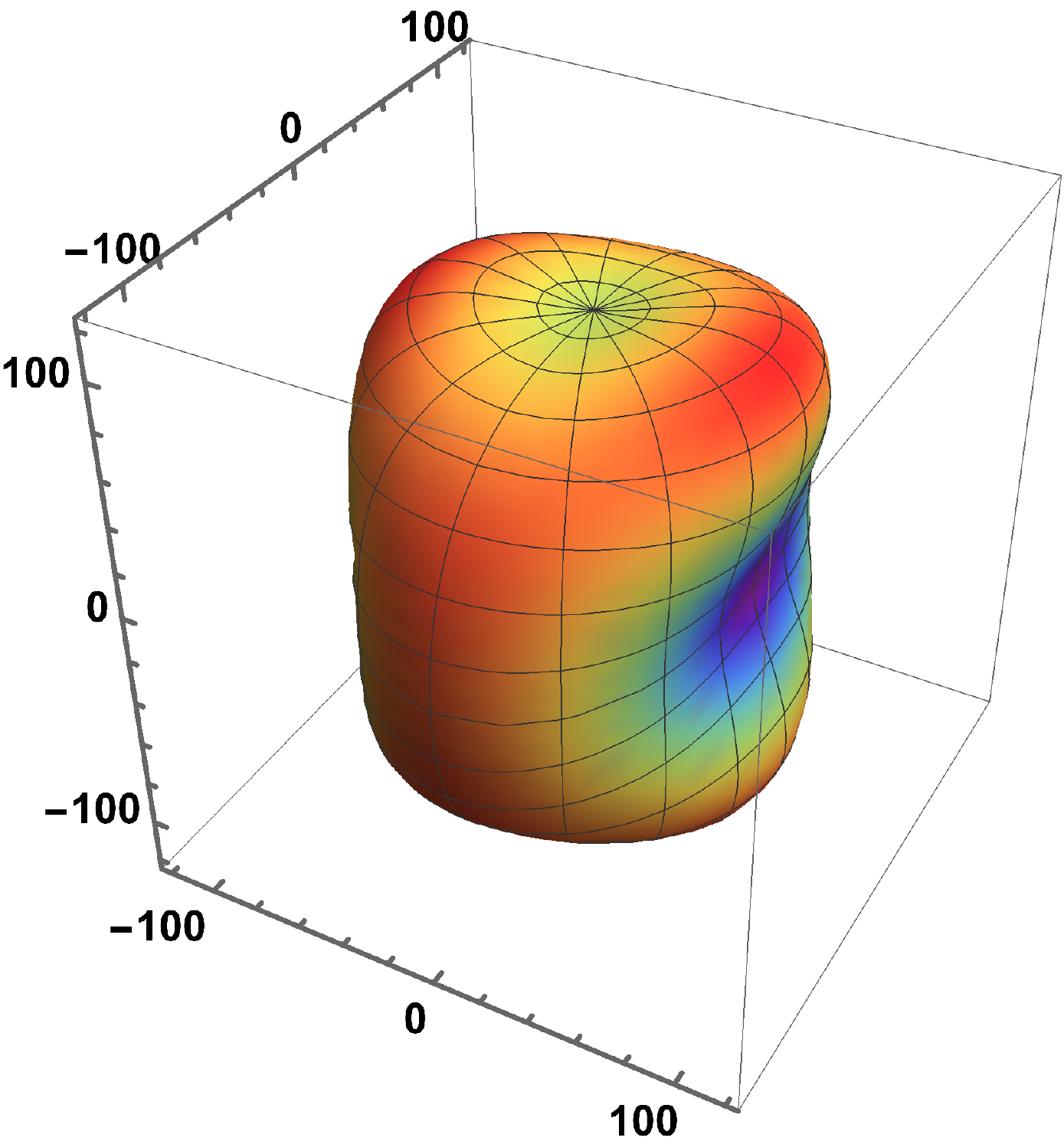}
         \includegraphics[height=0.2\textwidth]{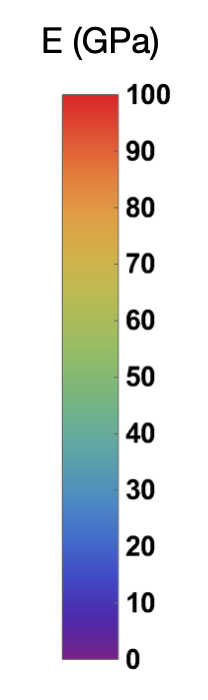}
    \caption{Directional dependence of Young's modulus of the martensite for different Ta content (18.75\%, 25\%, 31.25\%, 37.5\% and 43.75\%Ta content from left to right respectively).\label{fig:dir_ym_mar}
    }
\end{figure*}
\begin{figure*}
    \centering
        \includegraphics[width=0.18\textwidth]{aus-1875.pdf}
        \includegraphics[width=0.18\textwidth]{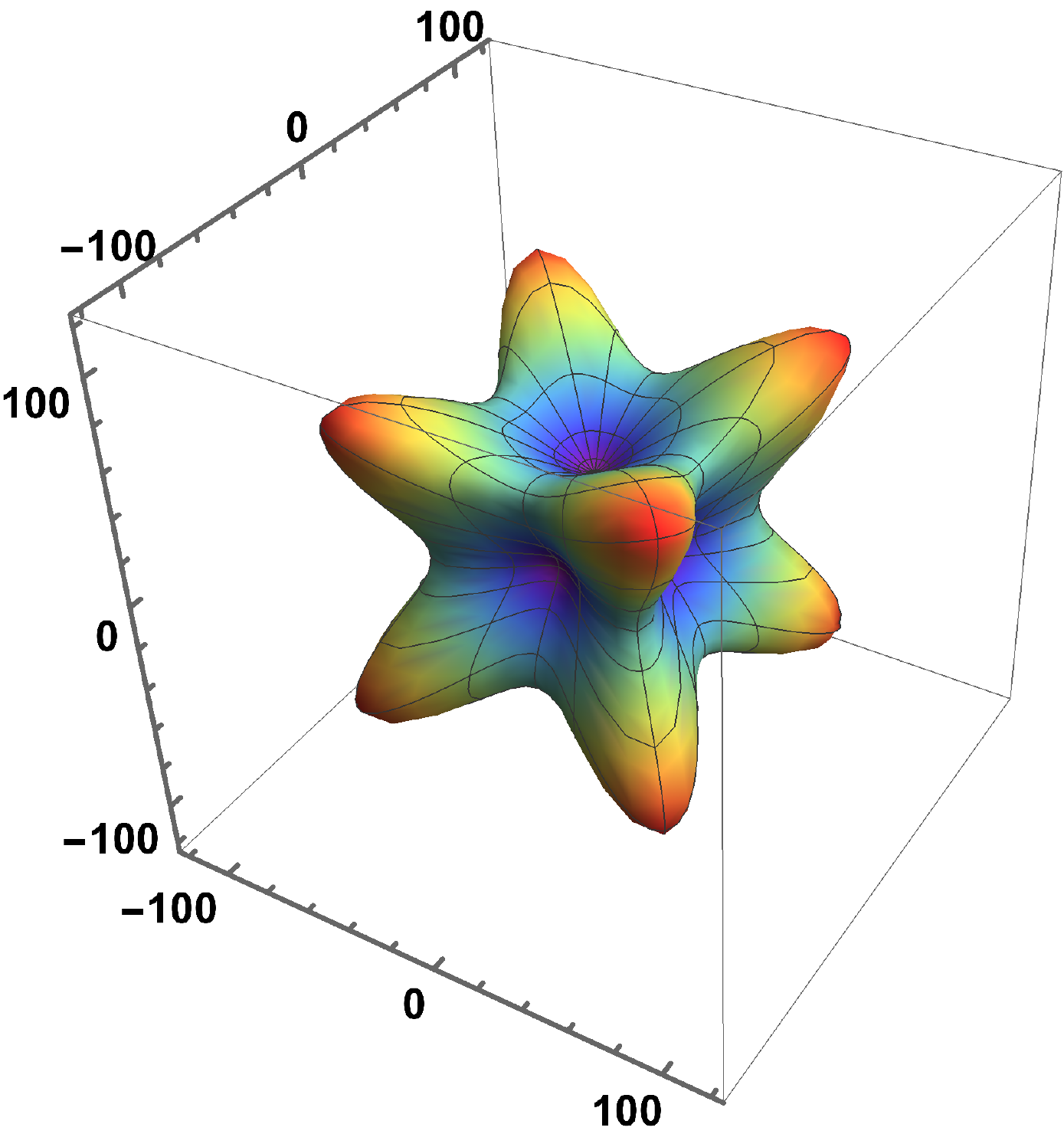}
        \includegraphics[width=0.18\textwidth]{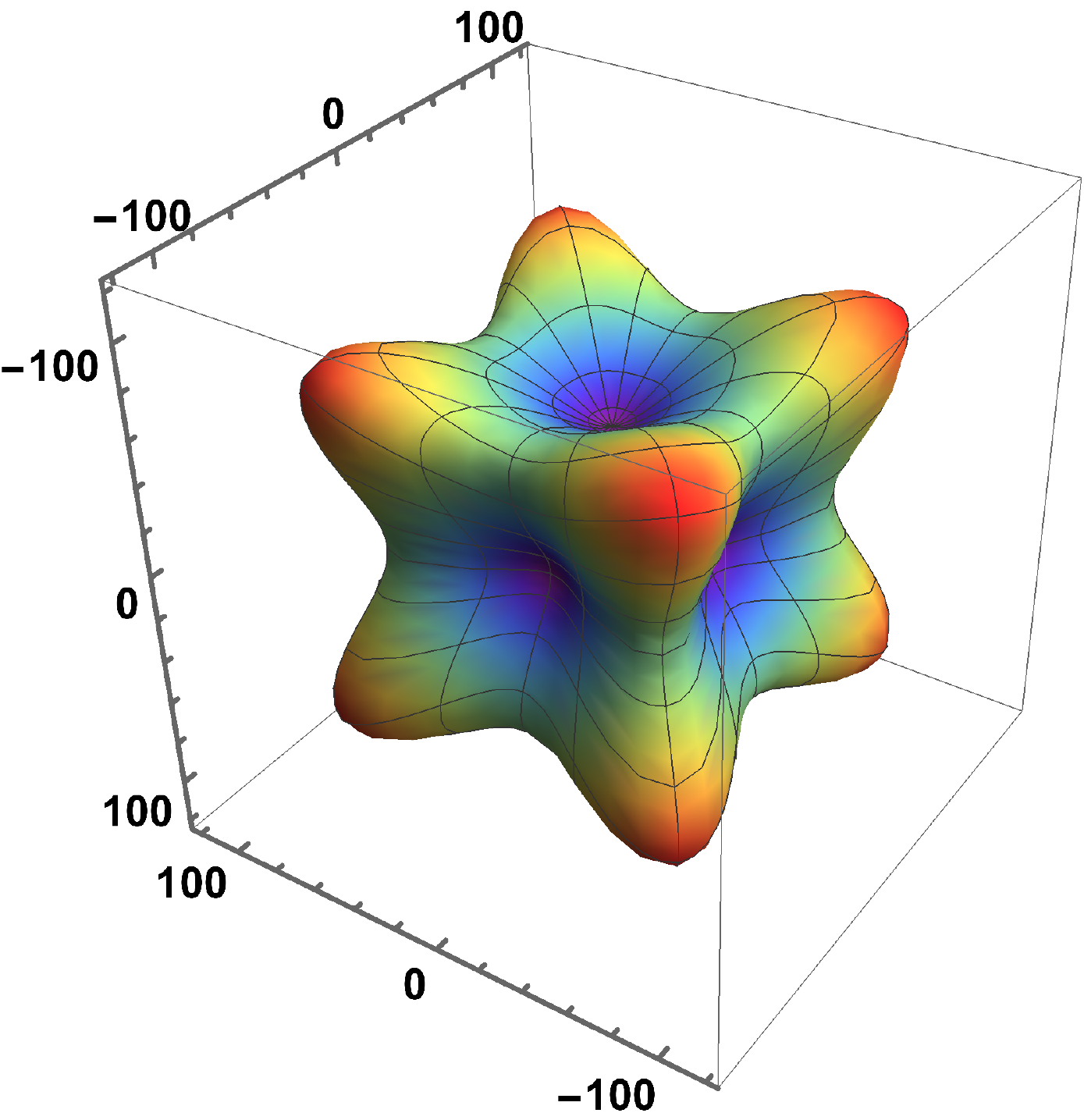}
        \includegraphics[width=0.18\textwidth]{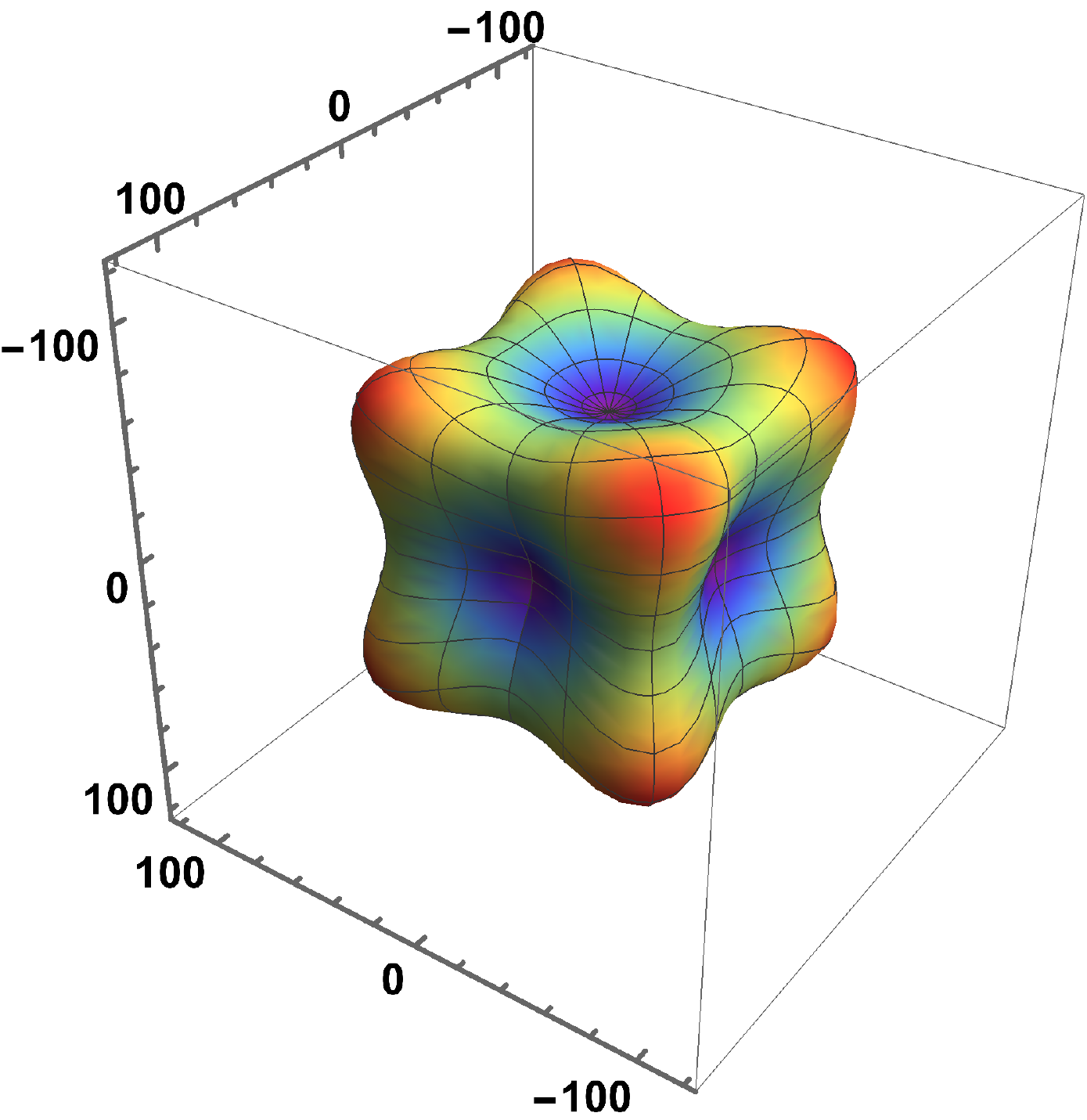}
        \includegraphics[width=0.18\textwidth]{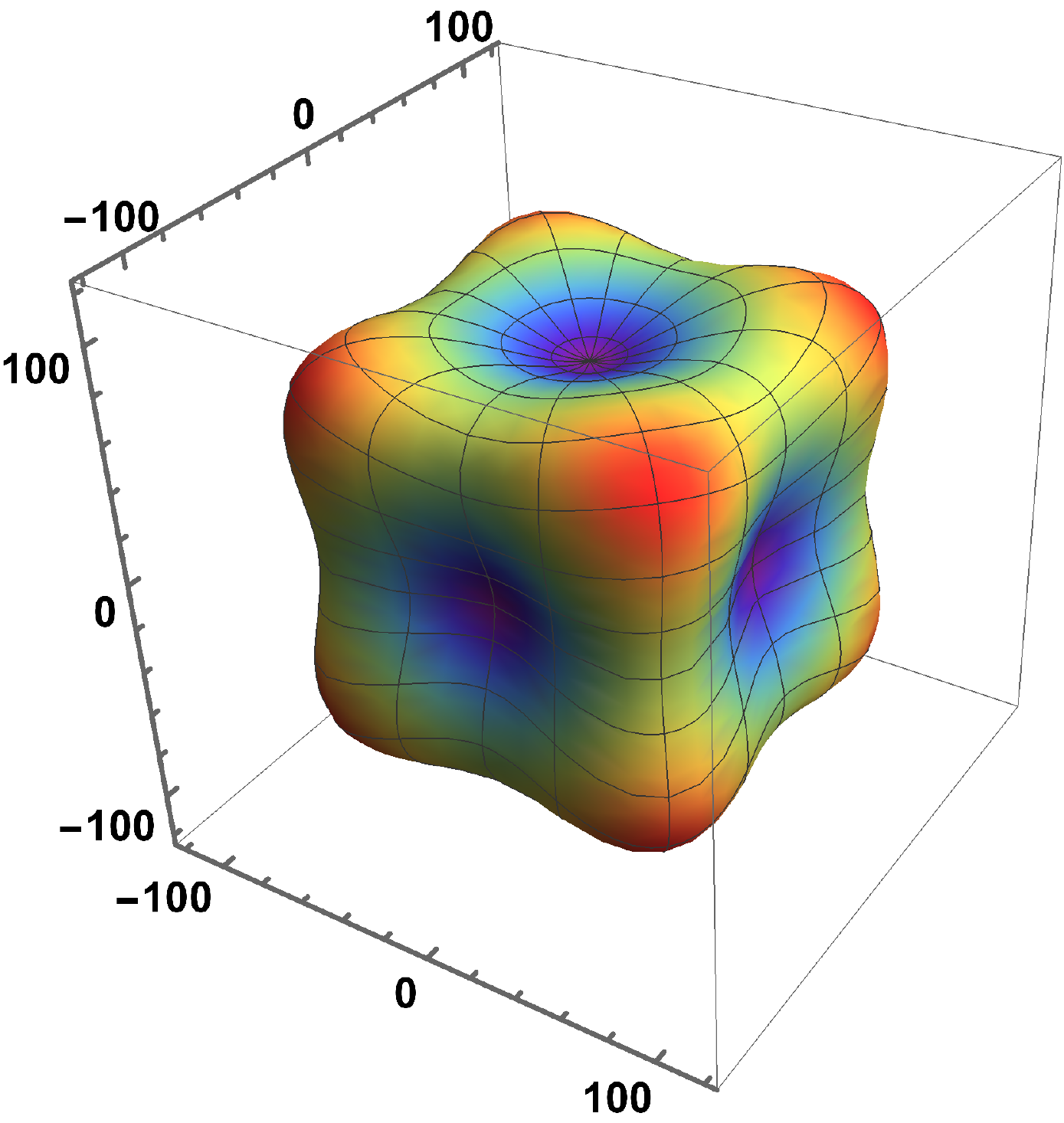}
        \includegraphics[height=0.2\textwidth]{bar.png}
    \caption{Directional dependence of Young's modulus of the austenite for different Ta content (18.75\%, 25\%, 31.25\%, 37.5\% and 43.75\%Ta content from left to right respectively).\label{fig:dir_ym_aus}
    }
\end{figure*}
\begin{equation}
 A^U = 5\frac{G_V}{G_R} + \frac{B_V}{B_R} - 6 \geq 0 \quad ,
\end{equation}
\begin{equation}
 A_C = \frac{G_V - G_R}{G_V + G_R} \quad .
\end{equation}
Both parameters have been extensively used in the materials community to study elastic anisotropy of materials~\cite{Xiao_109, Ravindran_84, Ganeshraj, Shu-Chun, Maxisch_73, Mao, Dong}. For a perfectly isotropic crystal $A^U$ =  $A^C$ = 0. There is no upper bound for $A^U$, whereas the maximum value of $A^C$ is 1. The deviation from 0, for both parameters, is a measure of the anisotropic behavior of a material. 

The calculated values of these two quantities for the austenite and martensite phases as a function of composition are shown in Fig.~\ref{fig:fig-5}. 
Both indices ($A^U$ and $A^C$) have large values for the two phases at low Ta concentrations suggesting a highly anisotropic nature. 
With increasing Ta content, $A^U$ and $A^C$ converge towards zero implying that the addition of Ta makes both phases more isotropic. 
The anisotropy appears to be less connected to the respective crystal phase, but more to the properties of the elements, since Ta is more isotropic than Ti.

The two, commonly used anisotropy indices are often not sufficient to fully depict the elastic behavior of different crystal phases. 
In order to further extend our description of the anisotropic behavior of the austenite and martensite phases we calculate the directional dependence of the Young's modulus as a function of composition. This approach has been well established for various systems~\cite{Xiao_109, Ravindran_84, Ma_245, Koval, Ghosh} to study the elastic anisotropy of materials.
The direction-dependent Young's modulus for the martensite (orthorhombic) is given by~\cite{Nye}
\begin{multline}
 1/E = l_1^{4}S_{11}+2l_1^{2}l_2^{2}S_{12}+2l_1^{2}l_3^{2}S_{13}+l_2^{4}S_{22}+2l_2^{2}l_3^{2}S_{23}+ \\ l_3^{4}S_{33}+l_2^{2}l_3^{2}S_{44}+l_1^{2}l_3^{2}S_{55}+l_1^{2}l_2^{2}S_{66} \quad .
 \label{eq:dir_ym}
 \end{multline}
Likewise, the direction-dependent Young's modulus for the austenite (cubic) is given by~\cite{Nye, Lee}
\begin{equation}
1/E_{hkl} = S_{11} - 2S_{A}\frac{h^{2}k^{2} + k^{2}l^{2} + l^{2}h^{2}}{h^{2} + k^{2} + l^{2}} \quad ,
\end{equation}
where $S_{A} = S_{11} - S_{12} - S_{44}/2$  and $S_{ij}$ are the elastic compliance matrix elements as introduced above.  $l_1$, $l_2$ and $l_3$ are the direction cosines, calculated using a spherical coordinate system i.e. $l_1 = \sin[\theta]\cos[\phi]$, $l_2 = \sin[\theta]\sin[\phi]$, and $l_3 = \cos[\theta]$. In Fig.~\ref{fig:dir_ym_mar} we show the directional dependence of the Young's modulus of the martensite as a function of composition. At low Ta content, the martensite phase reveals a strong anisotropy in the Young's modulus  that decreases smoothly with increasing Ta content, similar to the trend in the anisotropy indices discussed above. Likewise, in Fig.~\ref{fig:dir_ym_aus} the directional dependence of the Young's modulus of the austenite phase is shown as a function composition. The outcome is similar to what we already observe for martensite and again this is good agreement with the findings from our anisotropy indices discussion.  Similar behavior was also  observed in $\beta$-type Ti-Nb alloys~\cite{Moreno, Ma_245} where it was found that at 18.75\% Nb content the $\beta$ phase shows strong anisotropy in the Young's modulus, but at 31.25\% Nb the anisotropy decreases drastically and the material becomes almost fully isotropic. The austenite phase exhibits the highest $E$ values along the [111] axes and reveals exceptionally low Young's modulus ($\sim$30 GPa) in the [100], [010] and [100] directions, similar to $\beta$ Ti-Nb alloys~\cite{Moreno, Ma_245}.  In the martensite the highest $E$ values are observed in the [101] direction which is parallel to the [111] direction in austenite. Similarly, the Young's modulus in [100] direction (parallel to [001] austenite) is lowest, and remains the lowest with increasing Ta content. The overall trend again indicates that the anisotropy is more strongly related to the properties of the elements (Ti and Ta) than to the crystallographic structure.

\section{Summary}\label{conclusions}
We have investigated trends in the elastic properties of the martensite and austenite phases in Ti-Ta alloys over a wide range of compositions. The tetragonal shear constant of the high-temperature austenite phase strongly stiffens with increasing Ta content, making the phase more stable, which is in agreement with our previous work and existing literature.  Below 12.5\% Ta, the austenite phase becomes unstable. The martensite phase exhibits and unexpected \emph{non-linear} trend in the shear elastic constants,  with a maximum  at $\sim$ 30\% Ta.

The polycrystalline bulk, shear, and Young's moduli of the austenite show a linear dependence on the composition, whereas the shear and Young's moduli of the martensite again behaves non-linear, similar to the single crystal shear elastic constants. This gives rise to different thermodynamic and mechanical properties observed in the two phases which may have potential applications in other sectors e.g. in the biomedical industry where materials with low Young's modulus are desirable. Using a universal brittle-ductile phase diagram reveals that the addition of Ta enhances the strength but reduces the ductility in these alloys. 

Furthermore, the anisotropy in the Young's modulus is strongly composition dependent for both phases.  With increasing Ta content the elastic properties become more isotropic independent of the phase which indicates that this originates mainly in the isotropic nature of Ta.  This information is of particular interest in the study of microcrack formation.

The class of Ti-Ta alloys is not only of importance as high-temperature shape memory alloys, but might also show potential for biomedical applications, as sensors or actuators.
The presented trends in the elastic properties of the martensite and austenite phases in Ti-Ta alloys can be helpful in the design of new materials and aid experimentalists in the identification of promising alloy compositions.

\section*{Supplementary Material}
See Supplementary Material for details on the structural description of the martensite and austenite phases, their elastic constants and polycrystalline values, a brittle-ductile phase diagram and values of the anisotroy indices.

\begin{acknowledgments}
The work presented in this paper has been financially supported by the Deutsche Forschungsgemeinschaft (DFG) within the research unit FOR 1766 (High Temperature Shape Memory Alloys, http://www.for1766.de, Project No. 200999873, sub-project 3).
The authors thank  Ralf Drautz and  Davide Sangiovanni for critical comments and helpful discussions.
\end{acknowledgments}

\section*{Data availability}
The data that support the findings of this study are available within the article and its Supplementary Material. Additional data are available from the corresponding author upon reasonable request.

\nocite{*}
\bibliography{aip}

\end{document}


\title{Trends in elastic properties of Ti-Ta alloys from first-principles calculations: Supplementary material}

\author{Tanmoy Chakraborty}
\email{tchakra7@jhu.edu.}
\affiliation{ 
Interdisciplinary Centre for Advanced Materials Simulation, Ruhr-Universit{\"a}t Bochum, 44780 Bochum, Germany
}
\affiliation{Department of Materials Science and Engineering, Johns Hopkins University, Baltimore, Maryland 21218, USA}
\author{Jutta Rogal}%
\affiliation{ 
Interdisciplinary Centre for Advanced Materials Simulation, Ruhr-Universit{\"a}t Bochum, 44780 Bochum, Germany
}%

\date{\today}

\maketitle

\section{Simulation cell setup}\label{str. des.}

\begin{figure*}[!hb]
\begin{center}
 \includegraphics[width=1.0\textwidth]{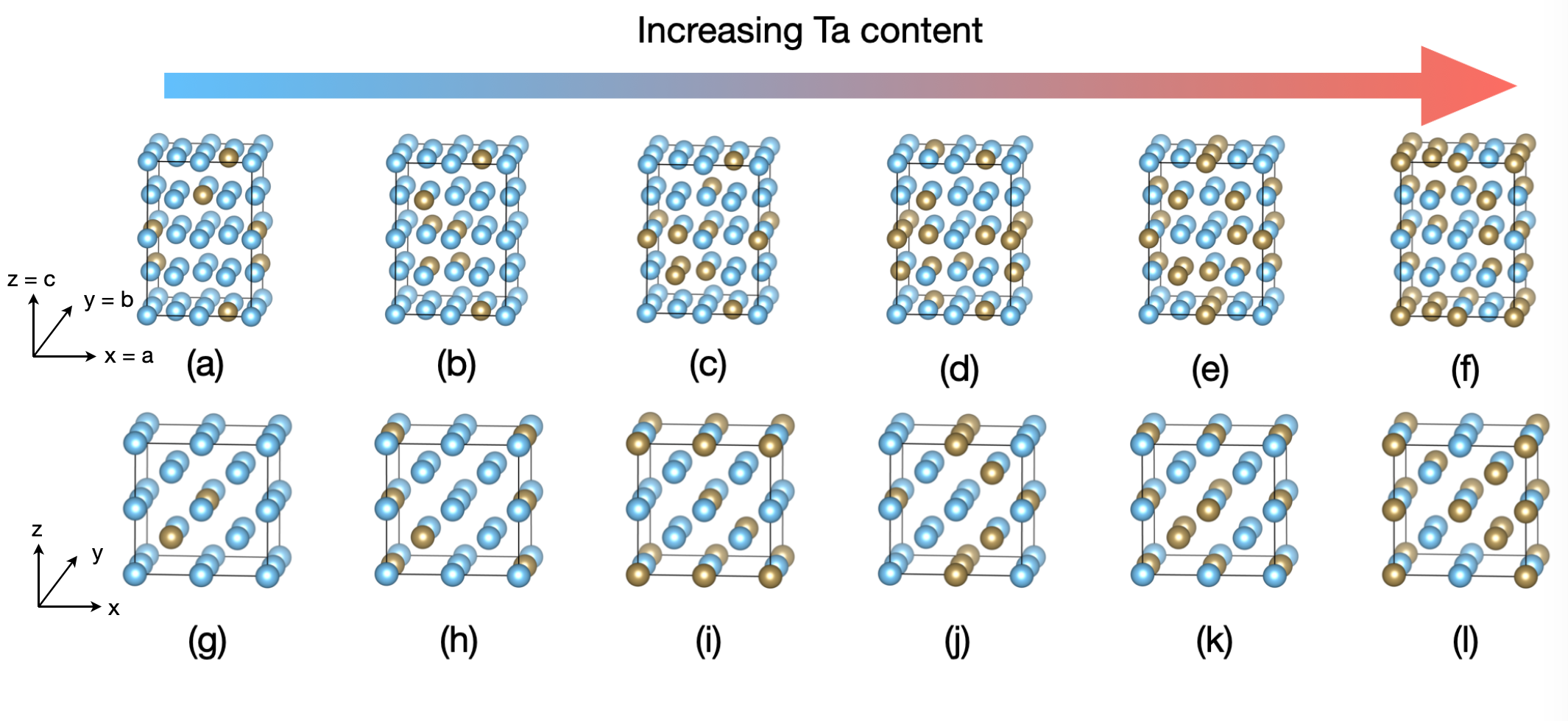}
\caption{\label{fig:fig-s1}
SQSs of the martensite (upper panel) and austenite (lower panel) phases as a function of increasing Ta content (from left $\to$ right).(a)-(f) denote Ti-12.5Ta, Ti-18.75Ta, Ti-25Ta, Ti-31.25Ta, Ti-37.5Ta, and Ti-43.75Ta martensite configurations, respectively. Likewise, (g)-(l) denote Ti-12.5Ta, Ti-18.75Ta, Ti-25Ta, Ti-31.25Ta, Ti-37.5Ta, and Ti-43.75Ta austenite configurations, respectively. Light blue and gold spheres represent the Ti and Ta atoms, respectively. The crystallographic directions are in accordance with Eq.~(15) in the main manuscript.
}
\end{center}
\end{figure*}

We have considered six different compositions of Ti-Ta alloys with 12.5\%, 18.75\%, 25\%, 31.25\%, 37.5\% and 43.75\% Ta. To model the chemical disorder of the martensite and austenite phases, we use special quasi-random structures (SQS)~\cite{Zunger} as implemented in the modified version~\cite{Pezold, Kossmann} of the ATAT package~\cite{Walle}. A (2$\times$2$\times2$) supercell is used to construct the SQSs of the martensite and austenite containing 32 and 16 atoms, respectively, for all compositions.  The values of the correlation function for the corresponding SQSs are 0.0165, 0.0105, 0.0045, 0.0044, 0.0045 and 0.004 for 12.5\%, 18.75\%, 25\%, 31.25\%, 37.5\%, and 43.75\%Ta for the martensite, respectively,  and 0.0722, 0.0439, 0.0236, 0.0121, 0.008 and 0.0752 for the austenite, respectively. A 7$\times$6$\times$6 and 12$\times$12$\times$12 $k$-point mesh were used for the martensite and austenite phase, respectively. In Fig.~\ref{fig:fig-s1} all composition dependent martensite and austenite SQS supercells used in the DFT calculations are shown.

\clearpage

\section{Elastic constants}

\begin{table}[!hb]
\caption{\label{pure_elastic} 
Elastic constants (in GPa) of bcc Ta and hcp Ti.}
 \begin{ruledtabular}
  \begin{tabular}{clcccccc}
  Element  &  Ref				& $C_{11}$ 	& $C_{12}$ 	& $C_{13}$ 	& $C_{33}$ 	& $C_{44}$	\\
   \hline\\
  Ta     	&  Present 			& 270.3 		& 162.4 		& 			& 			& 74.5  	\\
		&  Theo.~\cite{Ikehata}	& 257.2 		& 156.1 		& 			& 			& 70.6  	\\
		&  Expt.~\cite{Allard}		& 260.9 		& 157.4 		& 			& 			& 81.8  	\\
  Ti     	&  Present 			& 182.0 		& 62.8 		& 83.7 		& 147.4 		& 59.8 		\\
		&  Theo.~\cite{Ikehata}	& 171.6 		& 86.6 		& 72.6 		& 190.6 		& 41.1 		\\
		&  Expt.~\cite{Allard}		& 162.4 		& 92.0 		& 69.0 		& 180.7 		& 46.7 		\\
  \end{tabular}
  \end{ruledtabular}
\end{table}

\begin{table*}[!hb]
\caption{\label{elas_aus}
Calculated elastic constants (in GPa) for austenite in Ti-Ta  alloys.}
 \begin{ruledtabular}
  \begin{tabular}{lccc}
  Comp.  			& $C_{11}$ 	& $C_{12}$  	& $C_{44}$ \\
   \hline\\
  Ti-12.5Ta			& 115.0 		& 114.6 		& 41.1 \\
  Ti-18.75Ta 		& 127.8 		& 117.9 		& 41.3 \\
  Ti-25Ta 			& 137.7 		& 120.9 		& 42.6 \\
  Ti-31.25Ta		& 151.4		& 122.7 		& 44.6 \\
  Ti-37.5Ta 		& 158.6 		& 123.8 		& 45.1 \\
  Ti-43.75Ta		& 174.0 		& 126.6 		& 45.3 \\
  \end{tabular}
\end{ruledtabular}
\end{table*}

\begin{table*}[!hb]
\caption{\label{elas_mar}
Calculated elastic constants (in GPa) for martensite in Ti-Ta  alloys.}
 \begin{ruledtabular}
  \begin{tabular}{lccccccccc}
  Comp.  		& $C_{11}$ 	& $C_{22}$ 	& $C_{33}$ 	& $C_{44}$ 	& $C_{55}$ 	& $C_{66}$ 	& $C_{12}$ 	& $C_{13}$ 	& $C_{23}$\\ 
   \hline\\
  Ti-12.5Ta 	& 120.6 		& 177.0 		& 160.4 		& 21.7 		& 34.9 		& 31.3 		& 94.2 		& 128.2 		& 58.2 \\
  Ti-18.75Ta 	& 131.3 		& 182.0 		& 174.2 		& 54.1 		& 48.7 		& 31.7 		& 100.0 		& 129.5 		& 60.9 \\
  Ti-25Ta 		& 146.0 		& 182.2 		& 183.9 		& 66.0 		& 51.3 		& 37.2 		& 101.8 		& 125.3 		& 82.3 \\
  Ti-31.25Ta 	& 155.5 		& 188.2 		& 193.0 		& 67.6 		& 50.2 		& 37.9 		& 101.9 		& 129.0 		& 87.0 \\
  Ti-37.5Ta		& 163.4 		& 190.1 		& 200.3 		& 66.2		& 47.6 		& 36.7 		& 106.6 		& 130.2 		& 92.4 \\
  Ti-43.75Ta 	& 168.5 		& 190.7 		& 205.7 		& 55.1 		& 41.2 		& 30.8		& 110.6 		& 133.8 		& 99.3 \\
  \end{tabular}
 \end{ruledtabular}
\end{table*}

\begin{table*}[!hb]
\caption{\label{elastic_anisotropy}
Calculated bulk modulus ($B$), shear modulus ($G$) and Young's modulus ($E$) in GPa for the austenite and martensite phases as a function of Ta content.
}
\begin{ruledtabular}
  \begin{tabular}{lcccccc} \\
    & \multicolumn{3}{c}{Austenite} & \multicolumn{3}{c}{Martensite} \\
    \cline{2-4} \cline{5-7}\\
 Comp.  			& $B$ 	& $G$ & $E$  & $B$    & $G$  & $E$ \\
   \hline\\
  Ti-12.5Ta			& 114.7  	& 12.6 & 36.5 & 113.1  & 17.1  & 49.0 \\
  Ti-18.75Ta		& 121.1  	& 18.6 & 53.1 & 120.5  & 30.1  & 83.5 \\ 
  Ti-25Ta			& 126.5  	& 22.5 & 63.8 & 125.1  & 39.6  & 107.4 \\
  Ti-31.25Ta 		& 132.2  	& 28.3 & 79.3 & 129.9  & 41.9  & 113.5 \\
  Ti-37.5Ta			& 135.4  	& 30.7 & 85.8 & 134.3  & 42.2  & 114.6 \\
  Ti-43.75Ta		& 141.8  	& 35.7 & 98.8 & 138.6  & 38.2  & 105.1 \\
  \end{tabular}
   \end{ruledtabular}
\end{table*}

\clearpage

\section{Brittle-ductile phase diagram}\label{brittle-ductile}

\begin{figure*}[!hb]
\begin{center}
 \includegraphics[width=0.48\textwidth]{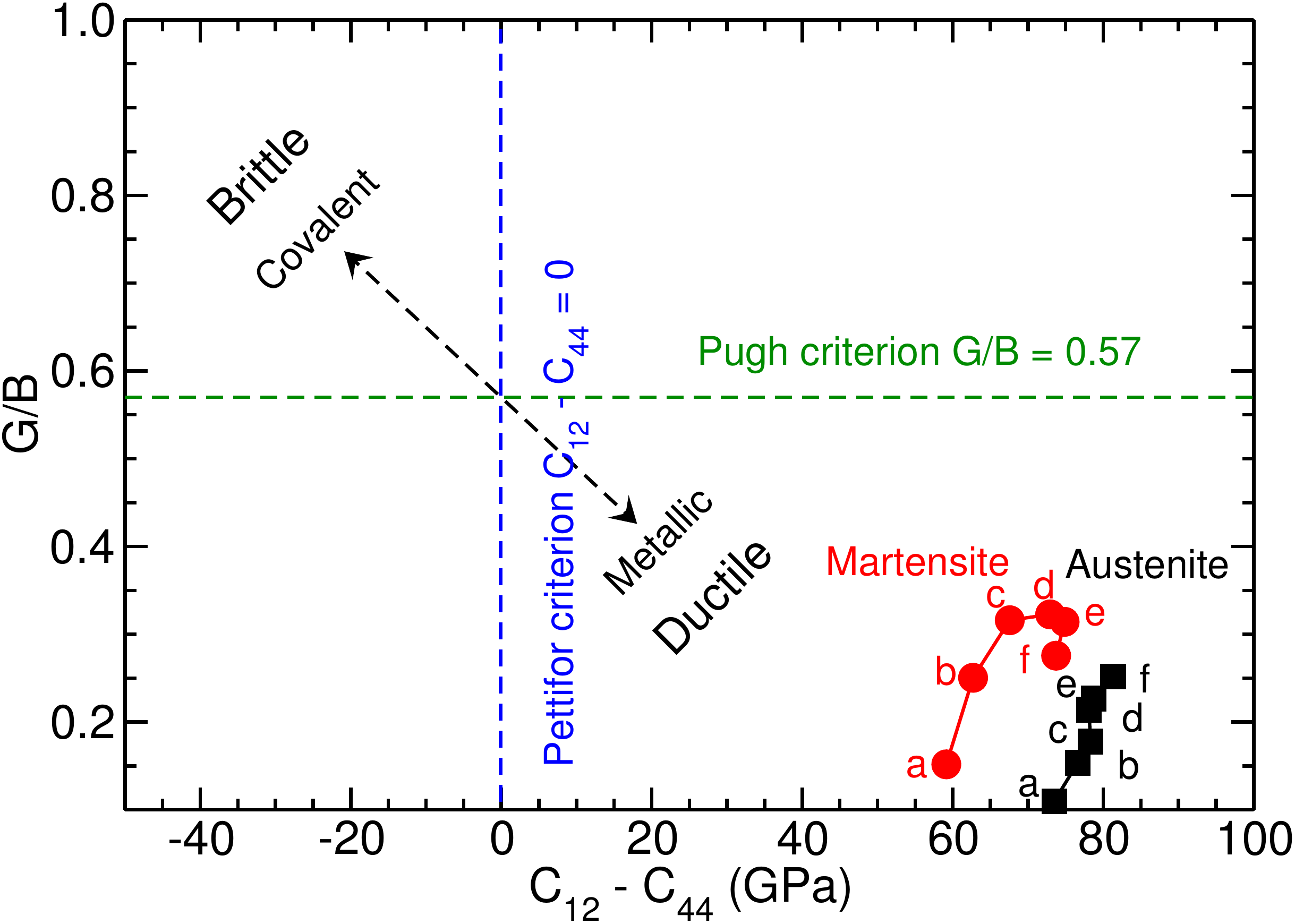}
 \caption{\label{fig:fig-s2}
 Brittle-ductile phase diagram of the martensite and austenite phases of Ti-Ta alloys according to Pugh and Pettifor criteria. a-f points represent 12.5\%, 18.75\%, 25\%, 31.25\%, 37.5\% and 43.75\% Ta content, respectively.
}
\end{center}
\end{figure*}

The ratio of shear modulus to bulk modulus ($G$/$B$) is a measure of the ductility/brittleness behavior of a material which was originally introduced by Pugh~\cite{Pugh} and is commonly known as Pugh's ratio in the materials community. Another parameter that is correlated with Pugh's ratio and is also often used to quantify the ductile/brittle nature of materials is known as Cauchy pressure~\cite{Pettifor}. A value of Pugh's ratio $> 0.57$ and a value of Cauchy pressure $< 0$ for a material is suggested to indicate that the material is having predominantly covalent character of the chemical bonds and consequently is predicted to behave brittle. We compute these two quantities for both phases over the entire composition range of interest in this work (see Fig.~\ref{fig:fig-s2}). We observe that with the addition of Ta both, Pugh's ratio and the Cauchy pressure, increase almost linearly for austenite, whereas for martensite initially both quantities first increase and then decrease. This behavior indicates that as  the Ta content is increased in Ti-Ta alloys, both phases become less ductile, however, they never become brittle at least not within the composition range studied in this work and, at the same time, they reveal metallic bonding. Therefore our computed values suggest that both, martensite and austenite, remain intrinsically ductile within the composition range studied in this work. Similar findings have also been reported for Ti-Nb alloys~\cite{Lai} where it was found that the bcc phase in Ti-Nb behaves as ductile over a wide range of Nb concentrations. This is again consistent with our Young's modulus results as higher Young's modulus indicate less ductility of the material and also with the Poisson's ratio discussed in our previous work~\cite{chakraborty_prb}. 

\clearpage

\section{Anisotropy indices}

\begin{table*}[!hb]
\caption{\label{anisotropy-index}
Calculated universal anisotropy index ($A^{U}$) and the percent anisotropy ($A_G$) of the austenite and martensite phases as a function of Ta content.
}
\begin{ruledtabular}
  \begin{tabular}{lcccc} \\
    & \multicolumn{2}{c}{Austenite} & \multicolumn{2}{c}{Martensite} \\
    \cline{2-3} \cline{4-5}\\
 Comp.  		& $A^{U}$ & $A_G$ & $A^{U}$ & $A_G$  \\
   \hline\\
  Ti-12.5Ta		& 44.2  & 0.96 & 24.7 & 0.71 \\
  Ti-18.75Ta	& 7.7    & 0.43 & 4.4   & 0.30 \\
  Ti-31.25Ta	& 3.9    & 0.28 & 1.3   & 0.12 \\
  Ti-31.25Ta	& 1.7    & 0.14 & 1.0   & 0.09 \\
  Ti-37.5Ta		& 1.1    & 0.10 & 0.7   & 0.06 \\
  Ti-43.75Ta	& 0.5    & 0.04 & 0.5   & 0.04 
  
  \end{tabular}
   \end{ruledtabular}
\end{table*}

\clearpage

\bibliography{aip}